\documentclass{webofc}
\usepackage[varg]{txfonts}   
\newcommand{\be}{\begin{equation}}
\newcommand{\ee}{\end{equation}}
\begin{document}
\title{Theory of hadronic molecules applied to the XYZ states}
%
%

\author{\firstname{Christoph} \lastname{Hanhart}\inst{1}\fnsep\thanks{\email{c.hanhart@fz-juelich.de}} 
}

\institute{Forschungszentrum J\"ulich, Institute for
           Advanced Simulation, Institut f\"ur Kernphysik and
           J\"ulich Center for Hadron Physics, D-52425 J\"ulich, Germany   }

\abstract{%
 In recent years data have been accumulated at various experiments about states in the
 heavy quarkonium mass range that seem to
 be inconsistent with the most simple variants of the quark model. 
 In this contribution it is demonstrated that most of those data are consistent with
 a dominant molecular nature of those states. It is also discussed which kind of
 observables are sensitive to the molecular component and which are not.}
\maketitle
\section{Introduction}
\label{intro}
Since 2003 various experiments all over the world such as Belle, BaBar, LHCb and BESIII, just to name only
the most relevant ones, have accumulated evidence for many new states which show properties that
appear to be inconsistent with the predictions of the until then very successful quark model.
All these states are located above the first open
heavy quark open--flavor threshold --- $\bar DD$ and $\bar BB$ for the charmonium and bottomonium
like systems, respectively. Furthermore, most of them have masses close to thresholds
of two narrow open--flavor states with quantum numbers consistent with these being in a relative $S$-waves. This 
makes those  good candidates for hadronic molecules --- bound states of the mentioned pair of
open--flavor states that need to be narrow for otherwise the composite systems would most 
probably be too broad to be observable~~\cite{Guo:2011dd,Filin:2010se}. For a recent review
of the physics of hadronic molecular states we refer to Ref.~\cite{Guo:2017jvc}
\section{Generalities}
\label{sec-1}
Hadronic molecules are bound states of two hadrons. In other words their wave function is dominated
by the two--hadron component. The probably most famous representative of this group of states is
the deuteron --- a proton--neutron bound state with a mass only 2.3 MeV smaller than the proton--neutron
threshold. There exist also light nuclei with even smaller binding energies --- e.g. the mass of the hypertriton that
may be envisioned as a bound state of the deuteron and a $\Lambda$ hyperon  is located only (130$\pm$ 50) keV
below the $d\Lambda$ break up threshold. It is therefore not the question, if hadronic molecules exist,
but, given the recent observations quoted in the introduction, if hadronic molecules formed by heavy open
flavor mesons exist.

If those composite states are located very close to an $S$--wave threshold they can be really 
large --- as a size estimate we may use $1/\gamma$, where $\gamma=\sqrt{2\mu E_b}$ denotes
the binding momentum~\footnote{The very same relation allows one to calculate from, e.g., the ground
state binding energy of hydrogen the corresponding Bohr-radius.} for $\mu$ and $E_b$ denoting 
the reduce mass of the constituents and their binding energy, respectively.  For example the size
of the deuteron turns out to be as large as 4.2 fm. If it is confirmed that the binding energy of $X(3872)$ --- a candidate
for a $D^0\bar D^{*\, 0}$ bound state --- is below 200 keV, its size would be above 14 fm. However,
it should be stressed that this size is a pure quantum effect~\footnote{It finds its origin in Heisenbergs uncertainty
principle: the constituents having an average momentum of $\gamma$ are to be found 
at a relative distance of $1/\gamma$ (throughout this text we use the convention that $\hbar=c=1$).}, which
is completely unrelated to the range of forces between the constituents which can be rather short --- for a detailed
discussion of the interplay of different scales in a bound state wave function
we refer the reader to Ref.~\cite{Hanhart:2007wa}. Furthermore,
the presence of a long ranged tale of the wave function that follows from the considerations above does not
prevent the existence of a short ranged component in the wave function --- however, this component
can not be quantified within the hadronic theory.

In order to approach the question, if hadronic molecules of heavy meson pairs exist,
 one first of all needs a well defined criterion that allows one to quantify
the molecular component in a given state. Such a criterion was put forward by S. Weinberg already more
than 50 years ago~\cite{Weinberg:1965zz}. He derived a formalism that allows one to 
relate the coupling strength of a given bound state  to the nearby continuum state, $g_{\rm eff}$, to the probability
to find this continuum state in the bound state wave function, $\lambda^2$. In particular he got for the situation
where the two constituents have the same mass
\be
\frac{g^2_{\rm eff}}{4\pi}=\frac{4M^2\gamma}{\mu}(1-\lambda^2) \ .
\ee
This coupling constant is nothing but the residue of the corresponding bound state pole and can
thus in principle be extracted from data. Moreover, up to corrections of order of $(R\gamma)$, where
$R$ denotes the range of forces, the scattering length and effective range for the scattering
of the two mesons can therefore be expressed in terms of $\lambda^2$ via
\be
a=-2\left(\frac{1-\lambda^2}{2-\lambda^2}\right)\frac{1}{\gamma} \ ; \quad r=-\left(\frac{\lambda^2}{1-\lambda^2}\right)\frac{1}{\gamma} \ .
\ee
While these scattering parameters are not experimentally accessible for a pair of heavy
mesons, they may be extracted from QCD lattice calculations by means of the so-called L\"uscher method~\cite{christopher}.

The original argument by Weinberg is applicable to stable bound states only. However, applying ideas
of Ref.~\cite{Bogdanova:1991zz} it was generalized to 
situations with remote thresholds in Ref.~\cite{Baru:2003qq} such that it may be applied to a large
number of $XYZ$ states --- for an alternative extension which, however, leads
to the same conclusion see Ref.~\cite{Gamermann:2009uq}. Moreover, the Weinberg criterion suggests that one may regard a prominence
of two--meson loop  effects in the properties of a given near threshold state that is driven by a large
coupling of the state to the two--meson component as an indication for a predominantly 
molecular state, even for cases where the $(R\gamma)$ corrections get sizable. An example 
of such a situation is the $Y(4260)$ that will be discussed in more detail below.
 
 It should be stressed that only those observables that are sensitive to the long ranged tail
 of the wave function of a molecular candidate qualify as being sensitive to their molecular
 component in the sense outlined above. All those observables sensitive to the short ranged
 parts of the wave function do not allow for any distinction between a molecular or a non--molecular
 state. Within the language of effective fields theories this translates into the fact that any
 observable that is sensitive to short ranged physics, must be accompanied by a local counter
 term at leading order preventing any quantitative predictions. Therefore in the following we
 distinguish observables sensitive to the short ranged components from those sensitive
 to the long ranged components.
 
\section{Features of the spectrum}
\label{sec-2}

Since hadronic molecules are expected first of all close to two--hadron thresholds
with quantum numbers derived from those potential constituents in a relative $S$-wave
one may deduce already a series of non-trivial statements about the molecular
spectrum directly from the location of those thresholds. To go beyond what is discussed
in this section detailed models guided by, e.g., the heavy quark spin symmetry
must be constructed as detailed in the contribution by V. Baru~\cite{vadim} presenting
the results of Refs.~\cite{Baru:2016iwj,Baru:2017gwo}.

\begin{figure}[t!]
\centering
\includegraphics[width=14cm,clip]{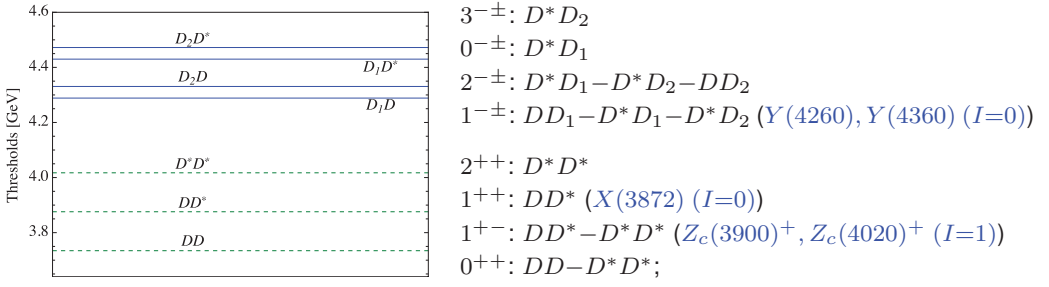}
\caption{Illustration of the location of the lowest $S$-wave thresholds of
narrow ope--flavor states together with a list of candidates of associated hadronic
molecules.}
\label{thresholds}       
\end{figure}

The lowest thresholds of relevance for molecular states with positive parity are
therefore $D\bar D$, $D\bar D^*$ and $D^*\bar D^*$ and for those with negative 
parity $D_1(2420)\bar D$, $D_1(2420)\bar D^*$, $D_2(2460)\bar D$ and $D_2(2460)\bar D^*$.
Their locations, shown in Fig.~\ref{thresholds}, already allow for a few non-trivial conclusions
on hadronic molecules. For example, one finds that the lightest negative parity state
should be $J^{PC}=1^{--}$ with
\be
M(1^{--})-M(1^{++})=388 \ \mbox{MeV} \simeq M(D_1(2420))-M(D^*) =410 \ \mbox{MeV}\ ,
\label{YXmassdiff}
\ee
where the vector state was identified with the $Y(4260)$ and the $1^{++}$ state with the $X(3872)$.
In particular it appears hardly possible to explain the $Y(4008)$ as a molecular structure. It is
therefore important to note that the newest BESIII data on the reaction $e^+e^-\to J/\psi \pi\pi$ does
not show any signal of the $Y(4008)$~\cite{Ablikim:2016qzw}.

In analogy to Eq.~(\ref{YXmassdiff}) one gets from the considerations above that
within the negative parity states 
\be
M(0^-)-M(1^-)\simeq M(D^*)-M(D)=140\ \mbox{MeV} \ ,
\ee
if the $0^-$ state exists. It is important to note that within the hadrocharmonium approach
the $0^-$ state is expected to be located by about the same mass difference below
the lowest $1^-$ state~\cite{Cleven:2015era}.

\section{Observables not sensitive to the molecular component: Radiative decays of $X(3872)$ and
$X$ production at high $p_T$}

An effective field theory working with hadronic degrees of freedom can not control hadronic wave functions
at short relative distances/large momenta. Accordingly, all observables sensitive to this momentum range
must therefore necessarily be described by an effective field theory that has a counter term already at
leading order. In other words: in this case the observable is not sensitive to the molecular component of 
the state. The effective field theory does not allow one to draw any conclusion of the physics underlying
the counter term --- this might be a $\bar QQ$ component, a tetraquark component or just the UV parts
of the heavy--meson loop. Prominent examples of observables that have a leading order counter term
are radiative decays of the $1^{++}$ $X(3872)$ into vector states. Accordingly, the ratio of branching
ratios of $X$ radiative decays into $\psi'$ and $J/\psi$ does not carry any information about the molecular
admixture in the wave function of $X(3872)$~\cite{Guo:2014taa} ---- contrary to an earlier claim of Ref.~\cite{Swanson:2004pp},
which was based on a particular model.

In complete analogy to the decays described above also the production of $X(3872)$ in, e.g., $pp$ or
$\bar pp$ collisions can be described only by employing a leading order counter term --- also here quantitatively
reliable predictions are not possible. However, it appears feasible to provide an order of magnitude 
estimate for the production rate. The estimate is based on the interesting observation made in
Ref.~\cite{Bignamini:2009sk} that (for concreteness formulated for a $\bar pp$ initial state --- the argument holds
equally well for $pp$ initial states)
\be
\sigma(\bar pp\to X+...)\leq \int_{\cal R} d^3k\left|\langle D^0\bar D^{* \ 0}(k)|\bar pp\rangle\right|^2 \ ,
\label{Xbound}
\ee
where the momentum cut-off $\cal R$ must be chosen large enough to saturate the bound state wave function.
In Ref.~\cite{Bignamini:2009sk}  $\cal R$ was identified with the binding momentum of the $X$ which lead
to an upper bound of the cross section various orders of magnitude below observations --- to find this bound
the right hand side of Eq.~(\ref{Xbound}) was estimated using Monte Carlo generators. However,
it is demonstrated in Ref.~\cite{Albaladejo:2017blx} on the example of the deuteron that in order to properly capture
the features of its well understood wave function values as large as twice to four times the pion mass
are appropriate for $\cal R$. Since the range of forces in the $X$ is expected to be the same as in the deuteron,
one can estimate
the subsequent rise in the rate estimate, which turns out to be sufficient to bring the bound of Eq.~(\ref{Xbound})
in line with the experimental rates. This conclusion is consistent with the much earlier studies of 
Refs.~\cite{Artoisenet:2009wk,Artoisenet:2010uu,Guo:2013ufa}.

\section{Observables sensitive to the molecular component: features of $Y(4260)$}

\begin{figure}[t!]
\centering
\includegraphics[width=12cm,clip]{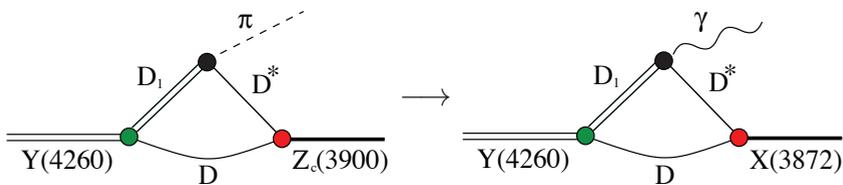}
\caption{Illustration of the intimate connection between the reactions $Y(4260)\to \pi Z_c(3900)$
and $Y(4260)\to \gamma X(3872)$ within the molecular picture for the $XYZ$ states.}
\label{Y2piZY2gX}       
\end{figure}

Contrary to the observables discussed in the previous section there are cases that are indeed sensitive to the
long ranged tail of the bound state wave function and to the best of our knowledge those are typically in line
with the predictions of the molecular model. An example is reported in Ref.~\cite{Guo:2013nza}, where
it was demonstrated that if $Y(4260)$ were a $D_1\bar D$ molecule with $J^{PC}=1^{--}$ and $Z_c(3900)$ as well as $X(3872)$
were $D\bar D^*$ molecules with $J^{PC}=1^{+-}$ and $J^{PC}=1^{++}$, respectively, then the observation of
the $Z_c(3900)$ in $Y(4260)\to Z_c(3900)\pi$ necessitates the reaction $Y(4260)\to X(3872)\gamma$, since
both transitions are based on the analogous production mechanism as illustrated in Fig.~\ref{Y2piZY2gX}.
This prediction was shortly after confirmed at BESIII~\cite{Ablikim:2013dyn}.
Note that also within the tetraquark picture, once adjusted to account for the spectrum of the isovector states,
both mentioned decays appear naturally~\cite{Maiani:2014aja}.

\begin{figure}
\centering
\sidecaption
\includegraphics[width=10cm,clip]{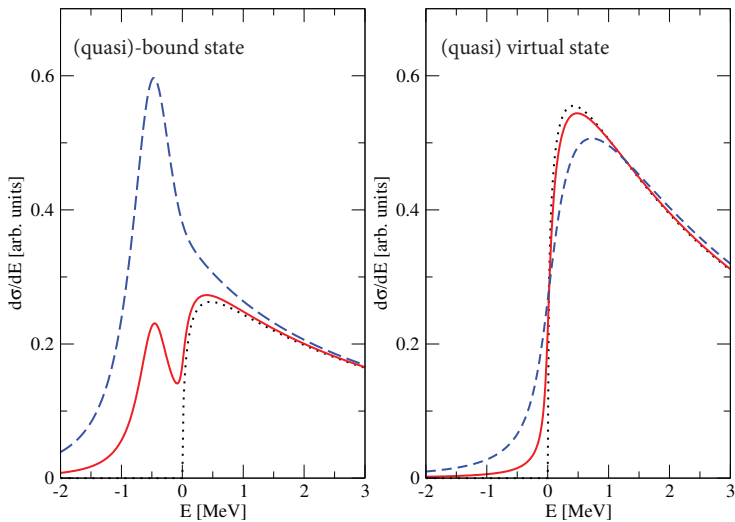}
\caption{Lineshapes that emerge naturally for molecular states with unstable constituents.
The curves were produced using the formulas of Ref.~\cite{Braaten:2007dw} for a pole located
0.5 MeV below the threshold on the first (second) sheet with respect to the studied channel.
For the dotted, solid and dashed lines a width of the constituent of 0, 0.1 and 1 MeV was assumed.
Furthermore an inelastic width of 1.5 MeV was introduced --- see Ref.~\cite{Guo:2017jvc} for details.}
\label{unstableconst}       
\end{figure}

An observable that is expected to provide most direct access to the molecular component of a given molecular
candidate is its signal in the elastic channel --- that is the channel which is assumed to form the bound state ---
since for a molecular state the coupling to this channel gets large. In Ref.~\cite{Braaten:2007dw} explicit expressions
are given that allow one to include straightforwardly the width of a constituent in the line shape of a molecular
state~\footnote{The equations of Ref.~\cite{Braaten:2007dw} acquire corrections for broader constituents that
are given in Ref.~\cite{Hanhart:2010wh}.}. The result of this is some leakage of strength of the amplitude below the nominal production threshold.
If this region of leakage overlaps with the pole position below threshold and if this pole is located on the physical
sheet with respect to that channel, a peak will develop in the distribution which will be the more pronounced the 
more overlap there is. If the pole is located on the unphysical sheet for the channel of the constituents no peak
will develop. 

Given the considerations above it becomes apparent that the preliminary BESIII data for $e^+e^-\to D^*\bar D\pi$~\cite{sun}
are consistent with a $Y(4260)$ that has a pole on the physical sheet for the $D_1\bar D$ channel and that
couples strongly to that channel. However, more detailed analyses are necessary before definite conclusions can
be drawn.

\section{Summary and outlook}

All states discovered since 2003 in the quarkonium sector that do not fit into the until
then very successful quark model are located above the lowest open--flavor threshold.
Many of them are candidates for hadronic molecules since they have masses close
to $S$--wave thresholds of pairs of narrow open--flavor states. 

In this presentation arguments based on fundamental concepts of effective field theories are presented
that show that various observables claimed to rule out a molecular interpretation of, e.g., the $X(3872)$
are in fact not sensitive to the molecular component at all. On the other hand it is demonstrated that
those observables that are largely saturated by the molecular component of potential molecular states
up to date are consistent with an interpretation of $X(3872)$ and $Z_c(3900)$ as $D\bar D^*$ molecules
and $Y(4260)$ as $D_1\bar D$ molecule. Analogous evidence can be put forward for various other
states --- for a recent review on hadronic molecules we refer to Ref.~\cite{Guo:2017jvc}.

In order to further consolidate this picture it appears necessary that more high quality data are being collected.
Of highest priority are here cross section measurements in the elastic channels (those are also
argued to distinguish poles of the $S$--matrix and cusp structures~\cite{Guo:2014iya}) as well
as data on additional states also for other quantum numbers as well as for the bottomonium 
sector. Clearly current experiments like LHCb and BESIII as well as future experiments such
as BelleII and PANDA are very well equipped to provide those data. I therefore expect a bright
and educating future for the spectroscopy of heavy mesons that will teach us a lot about the inner
workings of QCD.

\section*{Acknowledgments}

The author is grateful for the various enlightening collaborations with Vadim Baru, Martin Cleven, Feng-Kun Guo, Yulia Kalashnikova, 
 Ulf-G. Mei\ss ner, Alexey
Nefediev, Qian Wang, 
Qiang Zhao and Bing-Song Zou that lead to the results presented here.
This work is supported in part by the DFG and the NSFC through funds provided to the Sino-German CRC 110 ``Symmetries
and the Emergence of Structure
in QCD''.

\end{document}